\documentclass[apex,twocolumn]{jjap3}
\usepackage{txfonts}
\usepackage{graphicx}
\usepackage{epsfig}

\title{Optimized spin-injection efficiency and spin MOSFET operation based on low-barrier ferromagnet/insulator/n-Si tunnel contact}
\author{Yang Yang$^{1}$, Zhenhua Wu$^{2}$, Wen Yang$^{3}$, Jun Li$^{1}$\thanks{E-mail: lijun@xmu.edu.cn}, Songyan Chen$^{1}$, Cheng Li$^{1}$}
\inst{$^{1}$ Department of Physics, Semiconductor Photonics Research Center, Xiamen University, Xiamen 361005, China \\
$^{2}$ Key Laboratory of Microelectronic Devices and Integrated Technology, Institute of Microelectronics, Chinese Academy of Sciences, 100029 Beijing, China \\
$^{3}$ Beijing Computational Science Research Center, Beijing 100089, China
}

\abst{We theoretically investigate the spin injection in different FM/I/n-Si tunnel contacts by using the lattice NEGF method. We find that the tunnel contacts with low barrier materials such as TiO$_2$ and Ta$_{2}$O$_{5}$, have much lower resistances than the conventional barrier materials, resulting in a wider and attainable optimum parameters window for improving the spin injection efficiency and MR ratio of a vertical spin MOSFET. Additionally, we find the spin asymmetry coefficient of TiO$_2$ tunnel contact has a negative value, while that of Ta$_{2}$O$_{5}$ contact can be tuned between positive and negative values, by changing the parameters.
}

\begin{document}
\maketitle

The spin degrees of freedom have caught the eyes of researchers due to they shed lights on the next-generation devices with novel charge-spin integrated functionalities\cite{Igor2004}. Realizing the spin-based electronics (spintronics) on silicon, i.e., the most prevailing material in semiconductor industry, has special significance because the established mature Si-technology could greatly facilitate the productions and massive applications of spintronic devices. Fortunately, silicon is also considered as an ideal host for spintronics, as it exhibits long spin lifetime and diffusion length\cite{JansenRev1}.
In the past decade, milestone progresses have been achieved in Si-based spintronics. Room temperature electrical spin injection in silicon through the ferromagnet/insulator/Si (FM/I/Si) tunnel contacts with Al$_{2}$O$_{3}$, SiO$_{2}$ and crystalline MgO as barriers were claimed to be observed\cite{Dash2009,SiO2RT2011,Suzuki2011}. The spin polarized signals were detected by local three-terminal (3T)\cite{Dash2009,SiO2RT2011}, non-local-four-terminal (NL-4T) Hanle measurements\cite{Suzuki2011}, and the spin transport in Si channel were demonstrated in spin metal-oxide-semiconductor field-effect transistor (spin MOSFET)\cite{Sasaki2014,Takayuki2015}. Nevertheless, there remains challenges on obtaining clear and reliable signals, as well as understanding the spin transport process in the FM/I/Si tunnel contacts. The local 3T Hanle signals were under severely debates since they were recently found to be dominated by the defect-states-assisted hopping\cite{Inoue2015}, rather than the spin accumulation in silicon. While the spin signals of NL-4T, spin MOSFETs were still very weak\cite{Sasaki2014,Takayuki2015}, implying further optimizations of FM/I/Si contacts are required for their practical usages in spintronic devices.

As pointed out by Fert et al., a noticeable spin signal by the spin injection from a ferromagnet into semiconductor can be observed only if the contact resistance is engineered into an optimum window\cite{Fert2001}: the contact resistance cannot be too low to overcome the conductivity mismatch\cite{Schmidt2000}, nor be too high to keep the electron dwell time shorter than the spin lifetime\cite{Fert2007}. Min et al. revealed that the resistances of conventional tunnel contacts are orders of magnitude higher than the optimum value, due to the formation of Schottky barrier\cite{Min2006}. Therefore, controlling the contact resistance in a relatively low value is very important for enhancing the spin signals. Graphene as a low resistance material has been demonstrated to be a good tunnel layer for the efficient spin injection into silicon\cite{Erve2012}. It is straightforward to expect other low barrier materials, such as TiO$_2$ and Ta$_{2}$O$_{5}$, could also be used as the low resistance tunnel barriers for improving the spin injection efficiency. These low barrier materials have the advantage that they are compatible with the established Si-technology. Plus, the thicknesses of them can be adjusted, which offers a freedom to tune the contact resistance and also suppress the formation of paramagnetic silicide\cite{JansenRev1}. However, the spin transport process in low barrier tunnel contact is much complicated, since both the Schottky barrier and the thermionic emission could take important roles. Therefore, a unified model which takes account of those effects is necessary for studying the spin transport of low barrier FM/I/Si tunnel contacts.

In this paper, we present a theoretical investigation of the spin injection in different FM/I/n-Si tunnel contacts by the non-equilibrium Green's function (NEGF) method\cite{Datta_book,WYang_GF_2017}. The transmission coefficient of band profiles with various tunnel and Schottky barrier are calculated by the lattice Green's function. And the thermionic emission process is taken into account by the temperature-dependent Fermi energy of n-Si and the Fermi-Dirac distributions. By using this method, the spin polarization (SP) of injected current, its parameters-dependence, and the magnetoresistance (MR) ratio of a vertical spin MOSFET\cite{Sugahara2005} are studied and discussed.

\begin{figure}[tbp]\label{Fig1}
 \includegraphics[width=0.82\columnwidth]{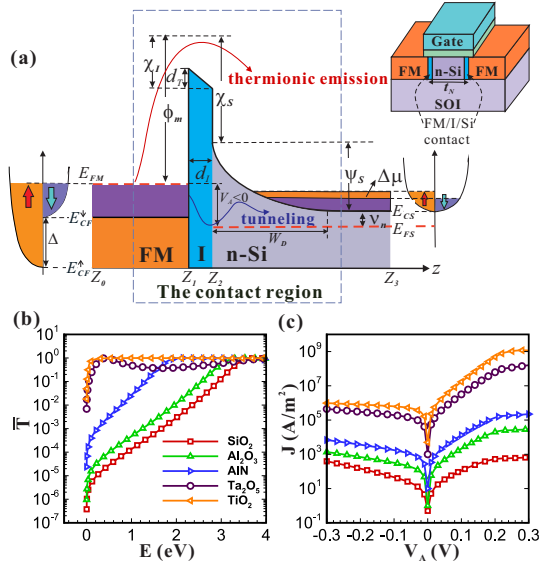}
 \caption{ (a) Schematic of the energy band profile of a FM/I/n-Si tunnel contact under a reverse bias ($V_{A}<0$) for the spin injection. The inset of (a) depicts the vertical spin MOSFET with a symmetric FM/I/n-Si/I/FM multilayer structure. (b) and (c) The typical results of the averaged transmission coefficient $\overline{T}$ ($V_{A}=0.2$ V and $k_{t}=0$) and the total current density $J$ ($d_{I}=1$ nm, $N_{D}=3\times 10^{18}$ cm$^{-3}$ and $T_{0}=300$ K), respectively.}
\end{figure}

The model of considered FM/I/n-Si tunnel contact is sketched in Fig. 1(a). The contact region $[Z_{0},Z_{3}]$ is assumed to be located in between two semi-infinite leads. $z \in [Z_{0},Z_{1}), [Z_{1},Z_{2}], (Z_{2},Z_{3}]$, corresponds to the ferromagnet, insulator barrier, n-Si, respectively. Similar to the two current model\cite{Mott}, the electrons with majority ($\uparrow$) and minority ($\downarrow$) spin can be viewed to flow in independent channels. In the contact region, the Hamiltonian operator for each spin channel is
\begin{equation} \label{eq1}
   \hat{H}_{C}^{\sigma}=-\frac{\hbar^{2}}{2}\frac{1}{\partial z}[\frac{1}{m^{*}_l(z)}\frac{\partial}{\partial z}]+\frac{\hbar^{2}k_{t}^{2}}{2m^{*}_t(z)}+U^{\sigma}(z),
\end{equation}
where $\sigma$[$\in (\uparrow, \downarrow)$] is the index of spin, $\hbar$ is the reduced Planck constant, $k_{t}$ is the transverse wave vector, and $m^{*}_{t(l)}(z)$ is the material-dependent transverse (longitudinal) electron effective mass. Here, we assume the electron effective mass $m^{*}_{F}$ ($m^{*}_{I}$) of ferromagnet (insulator) is isotropic, while the electron effective mass of n-Si is anisotropic, and $m^{*}_{St(l)}$ is the transverse (longitudinal) electron effective mass of n-Si.

$U^{\sigma}(z)$ denotes the potential energy profile function for the $\sigma$ spin channel, and consists of two parts
\begin{equation} \label{eq2}
   U^{\sigma}(z)=U_{CBO}^{\sigma}(z)+U_{S}(z).
\end{equation}
$U_{CBO}^{\sigma}(z)$ describes the profile of conduction band offset for $\sigma$ spin, and is dependent on the conduction band bottom of ferromagnet $E_{CF}^{\sigma}$, and the electron affinity  $\chi_{I}$ ($\chi_{S}$) of insulator barrier (n-Si). Due to the exchange interaction, $E_{CF}^{\uparrow(\downarrow)}$ is split by the exchange splitting energy $\Delta$. $U_{S}(z)$ describes the spin-independent Schottky barrier energy profile, which is induced by the charge accumulation nearby the FM/I and I/n-Si interfaces. Using the standard depletion layer approximation\cite{Donald_book}, $U_{S}(z)$ is determined by the work function of the ferromagnet $\phi_{m}$, the electron affinity $\chi_{S}$ of n-Si, the doping density $N_{D}$, the thickness of the insulator barrier $d_{I}$, the permittivity of insulator (n-Si) $\epsilon_{I}$ ($\epsilon_{S}$), and the Fermi energy $E_{FS}$ of n-Si. For n-Si from nondegenerate to degenerate regime, $E_{FS}$ is dependent on the doping density $N_{D}$ and the temperature $T_0$, and can be obtained by numerically solving the charge-neutral condition function\cite{Donald_book}.
At thermal equilibrium, the ferromagnet's Fermi energy $E_{FM}$ is equal to $E_{FS}$. If the contact is under an applied bias $V_{A}$, then $E_{FS}=E_{FM}+qV_{A}$, where $q$ is the elementary charge of electron. Note that for a reverse (forward) bias, i.e., $V_{A}<0$ ($V_{A}>0$) , the tunnel contact is in the spin injection (extraction) mode, respectively.

By discretizing the contact region into an uniformly spaced 1D grid with the spacing $a$, the Hamiltonian operator $\hat{H}_{C}^{\sigma}$ can be transformed into a $N\times N$ tridiagonal matrix $\mathbf{H_{C}^{\sigma}}$ by the method of finite differences\cite{Datta_book}, where $N$ is the total number of grid points. The retarded Green's function in the lattice representation can be expressed as follow
\begin{equation} \label{eq4}
\mathbf{G_{C}^{\sigma}}=[(E+i\eta)\mathbf{I}-\mathbf{H_{C}^{\sigma}}-\mathbf{\Sigma^{\sigma}_{L}}-\mathbf{\Sigma^{\sigma}_{R}}]^{-1},
\end{equation}
where $E$ is the electron transmission energy, $\mathbf{I}$ is the identity matrix and $\eta$ is an infinitesimally small positive number. The coupling of the contact to the left (right) semi-infinite lead is taken into account by a $N\times N$ matrix of self-energy $\mathbf{\Sigma^{\sigma}_{L(R)}}$
\begin{equation}\label{eq5}
\begin{aligned}
& \mathbf{\Sigma^{\sigma}_{L(R)}}=
&\left\{
\begin{array}{ll}
\Sigma_{ij}= -t_{i}e^{ik^{\sigma}_{L(R)}a}, \quad \mathrm{for} \quad i=j=1(N) \\
\Sigma_{ij}= 0 ,  \quad \quad \quad  \mathrm{otherwise} \\
\end{array}
\right.
\end{aligned}
\end{equation}
where $ k^{\sigma}_{L(R)}=\sqrt{2m^{*}_l(z_{0(N+1)})[E-E_{t}(z_{0(N+1)})-U^\sigma(z_{0(N+1)})]}/\hbar$ is the longitudinal wave vector of a electron with $\sigma$ spin in the left (right) lead, $E_t(z)=\hbar^2 k_t^2/[2 m_t^*(z)]$   $k^{\sigma}_{L(R)}$ is the transverse kinetic energy of electron, and $t_{n}=\hbar^{2}/[2m_{l}^{*}(z_{n})a^{2}]$ is the coupling strength between the nearest grid points. In the above expressions, $z_{n}$ denotes the coordinate of the n-th grid point. $z_0$ and $z_{N+1}$ are the coordinates of the first point in the left and right leads, respectively.
The transmission coefficient of $\sigma$ spin channel can be given by the NEGF formalism\cite{WYang_GF_2017} as
 \begin{equation} \label{eq6}
 {T}^{\sigma}(k_{t},E)=Trace[\mathbf{\Gamma^{\sigma}_{L}}\mathbf{G_{C}^{\sigma}}\mathbf{\Gamma^{\sigma}_{R}}\mathbf{G_{C}^{\sigma +}}],
 \end{equation}
 where $\mathbf{\Gamma^{\sigma}_{L(R)}} \equiv i[\mathbf{\Sigma^{\sigma}_{L(R)}}-\mathbf{\Sigma^{\sigma\dagger}_{L(R)}}]$ is the broadening matrix.
 The current density of $\sigma$ spin channel is then calculated by the Landauer formula\cite{Datta_book}
\begin{equation} \label{eq9}
J^{\sigma}=-\frac{q}{4\pi^{2}\hbar}\int^{\infty}_{0}\int^{\infty}_{0}T^{\sigma}(k_{t},E)[f_{L}(E)-f_{R}(E)]k_{t}dk_{t}dE,
\end{equation}
where $f_{L(R)}(E)\equiv 1/[e^{(E-E_{FM(FS)})/k_{B}T_0 }+1]$ is the Fermi-Dirac distribution
functions in the left (right) lead and $k_{B}$ is the Boltzmann constant.

For a contact with potential profile consisting of insulator barrier band offset and space-varying Schottky barrier, $T^{\sigma}$ and $J^{\sigma}$  can be calculated by Eq. (\ref{eq6}) and Eq. (\ref{eq9}). For the low tunnel barriers, a portion of free electrons could be thermally exited [determined by $f_{L(R)}(E)$], even to obtain higher energies over the barrier, so that $T^{\sigma}$ is close to 1. The transport of these electrons is not by tunnelling, but by the thermionic emission, and is automatically taken into account in this model. The typical results of the averaged transmission coefficient of the two spin channels, i.e., $\bar{T}\equiv (T^{\uparrow}+T^{\downarrow})/2$, and the total electric current density, i.e., $J \equiv J^{\uparrow} + J^{\downarrow}$, are displayed in Fig. 1(b) and (c). We can see the exponentially varying feature of $\bar{T}$, and the current rectifying effect of Schottky contact are well reproduced by our calculations. Note that, for contact in the spin extraction mode, though the calculated SP is found to be 25-60\% smaller, the spin injection efficiency is generally higher than in the spin injection mode. Because for $V_A > 0$, the depletion region is suppressed, and the contact resistance can be lowered by orders of magnitude so that the spin depolarization can be alleviated. In the following, we will focus on the tunnel contacts in spin extraction mode, e.g., for $V_{A}=+0.2$ V.

\begin{table}[tpb]
\caption{Parameters of different insulator materials for the tunnel barriers.}
\setlength{\tabcolsep}{2pt}
\begin{tabular}[c]{cccccc} 
\Hline
& SiO$_{2}$\cite{Yee-Chia2003}   & Al$_{2}$O$_{3}$\cite{Yee-Chia2003}  & AlN (MgO)\cite{Sean2015,note1}  & Ta$_{2}$O$_{5}$\cite{Casperson2002}  & TiO$_{2}$\cite{Casperson2002} \\
\Hline
$\chi_{I}-\chi_{S} \ (eV) $ \  & 3.1  & 2.8  & 1.6 (1.5)  & 0.3  & 0 \\
$m_{I}^{*} \ (m_{e}) $   & 0.4    & 0.35   & 0.33 (0.35)   & 0.1  & 1.0 \\
$\epsilon_{I} \ (\epsilon_{0})$   & 3.9   & 10  & 8.5   & 25  & 31 \\
\Hline
\end{tabular}  \label{tab:parameters}
\end{table}

In the calculations, we have assumed the contact is grown along the [001]-orientation of silicon, so $m^*_{St(l)} =$  0.20 (0.92) $m_{e}$ (at 300 K), $\epsilon_{s}=$ 11.5 $\epsilon_{0}$ and  $\chi_{S} =$ 4.2 eV, where $m_{e}$ and $\epsilon_{0}$ is the free electron mass and the permittivity of vacuum, respectively. The ferromagnet material is chosen to be Fe, of which the parameters are $\phi_{m}=4.5$ eV , $m^*_F=2.3$ $m_e$ and $\Delta=1.5$ eV. These parameters could recover the Fermi wave vector $k^{\uparrow (\downarrow)}_F =$  1.05 (0.44) {\AA}$^{-1}$ for the $\uparrow$ ($\downarrow$) spin of Fe\cite{Grundler2001}. The parameters for different insulator barriers are listed in TABLE \ref{tab:parameters}. The effective RA product $r_{b}^{*}$ of tunnel contact, and the spin asymmetry coefficient (of contact resistance) $\gamma$ is defined by
\begin{equation} \label{eq11}
 r_{\uparrow(\downarrow)}=2r_{b}^{*}[1-(+)\gamma],
\end{equation}
where $r_{\sigma}\equiv V_{A}/J^{\sigma}$ is the individual RA product for $\sigma$ spin.
If the electron travels beyond the ballistic regime, the spin accumulation should be described by the spin drift-diffusion model\cite{Schmidt2000}.
Following Fert's derivation, the SP of injected current in silicon can be given by Eq. (20) of Ref. 9.

\begin{figure}[tbp]\label{Fig2}
 \includegraphics[width=0.85\columnwidth]{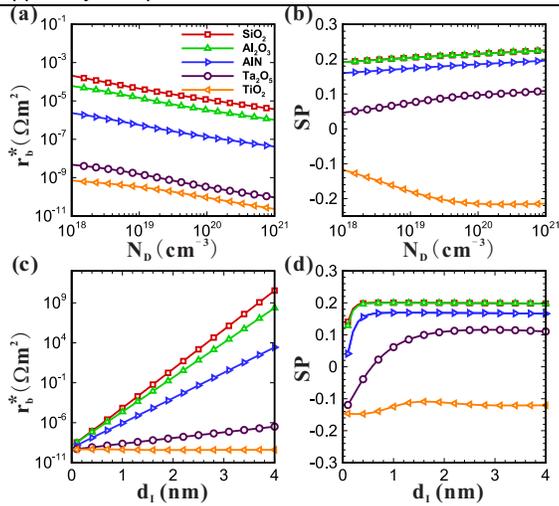}
 \caption{(a)-(b) Dependence of $r_{b}^{*}$ and SP as a function of the doping density $N_{D}$ of n-Si for FM/I/n-Si contact with different insulator barriers ($d_{I}=1$ nm).
 (c)-(d) The same as (a)-(b), but as a function of the thickness of barrier $d_{I}$ ($N_{D}=5\times 10^{18}$ cm$^{-3}$). $V_{A}=0.2$ V (in spin extraction mode) and  $T_{0}=300$ K is assumed in this figure.}
\end{figure}

In Fig. 2(a) and (c), we show the calculation results of $r_{b}^{*}$ as a function of the doping density $N_{D}$ and the thickness of barrier $d_{I}$.
As expected, for a 1-nm-thick barrier, $r_{b}^{*}$ can be reduced up to 5 orders by changing the barrier material from SiO$_2$ to TiO$_2$, i.e., decreasing the barrier height. While by increasing $N_{D}$ from $10^{18}$ to $10^{21}$ cm$^{-3}$, $r_{b}^{*}$ can only be adjusted by less than 3 orders. Though $r_{b}^{*}$ is very sensitively dependent on $d_{I}$ (except TiO$_2$ with a 0 eV barrier height) [see in Fig. 2(c)], the optimum value of $r_{b}^{*}$ ($\approx 10^{-8}$ $\Omega \cdot$m$^2$)\cite{Fert2001} for a noticeable MR requires an ultrathin layer with $d_{I}<0.5$ nm for the conventional tunnel barriers as SiO$_2$, Al$_2$O$_3$ and AlN. Experimentally, there are considerable difficulties in fabricating a sub-nanometer-thick layer with high qualities, i.e., with uniform and planar interfaces, very few defects and trapped-charges. Besides, the paramagnetic silicide, which is harmful to the spin transport\cite{JansenRev1}, can hardly be prevented from forming by such a thin layer ($d_{I}<0.5$ nm). Therefore, we can see the low barrier materials, like TiO$_2$ and Ta$_2$O$_5$, could offer wider range of tunable thickness\cite{note2} to balance the required resistances and the contact qualities.

 For SiO$_2$, Al$_2$O$_3$ and AlN contacts with $d_{I}>$ 0.5 nm, the SP of injected current are almost independent of $N_{D}$ and $d_{I}$ [see in Fig. 2(b) and (d)]. The reason is that $r_{b}^{*}$ of these contacts are much larger than the spin resistance $r_N$ ($\equiv \rho_{N}l_{sf}^{N}$) of n-Si, resulting in SP to be saturated at $\mathrm{SP}=\gamma \approx$ 0.17-0.2\cite{Fert2001}. For low barriers tunnel contacts, the behaviour of SP versus $N_{D}$ and $d_{I}$ are very different. We observe the SP of TiO$_2$ contact has a negative value, namely, the polarization direction of injected spins in silicon is opposite to that in ferromagnet. This is because the minority spin in ferromagnet has a smaller $k^{\downarrow}_F$ (than $k^{\uparrow}_F$), which better matches the relatively small evanescent wave vector ($\propto \sqrt{\chi_{I}-\chi_{S}-E}$) inside the TiO$_2$ barrier, leading to a larger $T^{\downarrow}$ than $T^{\uparrow}$\cite{Valenzuela2005}. Thus for TiO$_2$ contact, $J^{\downarrow}$ is larger than $J^{\uparrow}$, and a negative $\gamma$ (or SP) is produced.
 In contrast, for conventional barriers, like SiO$_2$, Al$_2$O$_3$ and AlN, the majority spin matches the the evanescent wave vector better, which makes $T^{\uparrow}$ larger than $T^{\downarrow}$, and $\gamma$ (or SP) be positive.

 \begin{figure}[tbp]\label{Fig3}
 \includegraphics[width=0.88\columnwidth]{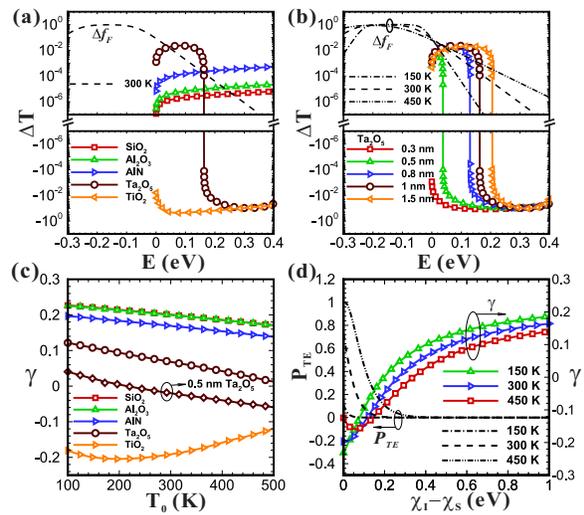}
 \caption{ (a) $\Delta T$ as a function of the electron transmission energy $E$, for different tunnel contacts ($V_{A}= 0.2$ V, $d_I = 1$ nm and $T_0 =$ 300 K). (b) The same as (a), but for Ta$_2$O$_5$ contact with different $d_I$. The black dash-dot, dashed, and dash-dot-dot lines depict $\Delta f_F (E)$ at 150, 300 and 450 K, respectively. (c) $\gamma$ as a function of temperature $T_0$ for different tunnel contacts. (d) $P_{TE}$ and $\gamma$ as a function of the tunnel barrier height $\chi_I-\chi_S$ ($m_{I}^{*}=$ 0.2 $m_e$ and $\epsilon_{I}=$ 11.5 $\epsilon_{0}$), at different temperatures. }
\end{figure}

 To demonstrate this, in Fig. 3(a) we plot the difference of transmission coefficient of the two spin channels, i.e, $\Delta T \equiv T^{\uparrow}-T^{\downarrow}$, as a function of the electron transmission energy $E$. The black dashed line of this figure denotes the difference of the Fermi-Dirac distribution functions, i.e., $\Delta f_F(E) \equiv f_{R}(E) - f_{L}(E)$, which determines the contribution of an electron with energy $E$ to the current. From $\Delta f_F(E)$, we can see the effective energy range is $0<E< 0.16$ eV. For $E$ out of this range, the contribution $\Delta f_F (E)$ falls below $10^{-4}$. Because in this range, $\Delta T$ of TiO${_2}$ is negative, $\gamma$ has a negative value. While for other barriers (with $d_I=1$ nm), $\Delta T$ in this range are positive, so $\gamma$ of them are positive too. Interestingly, we find the $\gamma$ (or SP) for Ta$_2$O$_5$ can be tuned from positive to negative by decreasing $d_{I}$, as exhibited in Fig. 2(d). The reason for the sign change of $\gamma$ can be illustrated in Fig. 3(b): the positive (negative) region of $\Delta T$ shrinks (expands) with decreasing $d_{I}$. Besides, by changing the temperature, the broadening of $\Delta f_F(E)$ can be varied, resulting in $\gamma$ for Ta$_2$O$_5$ be more sensitively dependent on the temperature, compared to other barriers [see in Fig. 3(c)]. For a 0.5-nm-thick Ta$_2$O$_5$ contact, $\gamma$ can even be tuned from positive to negative by increasing temperature. While for TiO$_2$, the $\gamma$ shows a non-monotonic dependence of temperature. The reason can be ascribed to the impact of thermionic emission transport, which is prominent in the low barrier contact. In Fig. 3(d), we plot the proportion of thermionic emission, i.e., $P_{TE}$, and $\gamma$ as a function of the tunnel barrier height $\chi_I-\chi_S$. At 300 K, we can see the thermionic emission takes place only for $\chi_I-\chi_S <$ 0.2 eV. With decreasing barrier height and/or increasing temperature, the proportion of thermionic emission increases. Also, we can see a crossover of $\gamma$ from negative to positive by increasing the barrier height, which is consistent with the preceding discussions, i.e., a low barrier contact could have a negative $\gamma$.

\begin{figure}[tbp]\label{Fig4}
 \includegraphics[width=0.90\columnwidth]{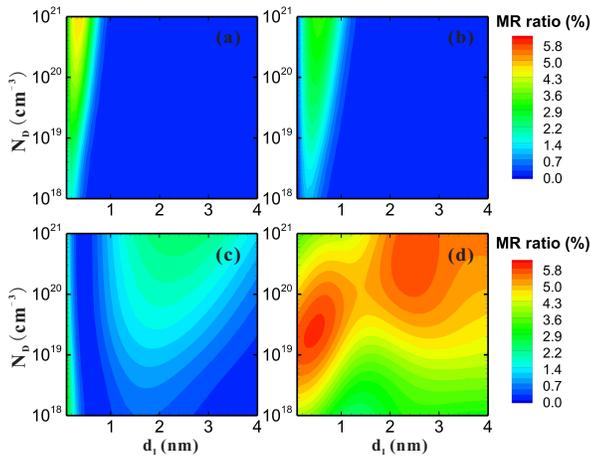}
 \caption{ Calculated MR ratio of a vertical spin MOSFET with symmetric FM/I/n-Si/I/FM structure as a function of $d_{I}$ and $N_{D}$. (a)-(d) the results of spin MOSFETs with Al$_2$O$_3$, AlN, Ta$_2$O$_5$ and TiO$_{2}$ barriers, respectively. $t_N =$ 100 nm and $T_{0}=300$ K is assumed in the calculations.}
\end{figure}

The vertical spin MOSFET can be modeled as a structure consisting of a symmetric FM/I/n-Si/I/FM multilayer\cite{Min2006,Sugahara2005} [see the inset of Fig 1(a)], of which the two terminal MR ratio can be calculated according to the analytical equations of Fert and Jaffr\`{e}s\cite{Fert2001}. In Fig. 4, we present the results of the MR ratio of vertical spin MOSFETs with a moderate channel length, i.e., $t_N =$ 100 nm. From panel (a) to (d), we can compare the optimum parameters windows for the MR ratio of spin MOSFETs with Al$_2$O$_3$, AlN, Ta$_2$O$_5$ and TiO$_2$ barriers.  For conventional barriers, such as Al$_2$O$_3$, AlN, a MR ratio $> 2\%$ usually requires a heavy doping with $N_D > 10^{20}$ cm$^{-3}$ , and an ultrathin barrier with $d_I < 1$ nm. These conditions are hard to achieve in experiments, and this conclusion is consistent with the obstacle revealed by the previous works\cite{Min2006,Erve2012}. In contrast, the magnitude of MR ratio and the optimum window is much larger for TiO$_2$. With $N_D = 6 \times 10^{18}$ cm$^{-3}$ (nondegenerate n-Si) and $d_I = 1$ nm, a MR ratio $\approx 4\%$ can be obtained by using TiO$_2$. For Ta$_2$O$_5$, the MR ratio might be suppressed in certain regions, such as $d_I$ in the range of 0.5 $\sim$ 1.2 nm. The reason is due to the sign change of $\gamma$ occurring in this region.
But a moderate value of MR ratio $\approx 2\%$ can still be obtained at a relatively large barrier thickness $d_I = 2.5$ nm. The optimized MR ratio and parameters window of spin MOSFETs with low barrier tunnel contacts could offer improved performance of Si-based spintronic devices.

In summary, we investigate theoretically the spin injection in the FM/I/n-Si tunnel contacts. We find that $r_{b}^{*}$ of contacts with low barriers, such as TiO$_2$ and Ta$_{2}$O$_{5}$, are orders of magnitude smaller than that of the conventional tunnel contacts. Therefore, the maximum MR signal and optimum parameters window for TiO$_2$ and Ta$_{2}$O$_{5}$ contacts are larger than the conventional tunnel contacts. Interestingly, we also demonstrate the spin asymmetry coefficient $\gamma$ of TiO$_2$ contact has a negative value, and $\gamma$ of Ta$_{2}$O$_{5}$ contact can be tuned from negative to positive by changing the thickness of tunnel barrier and temperature. The optimized spin signals and unique spin asymmetry properties of low barrier tunnel contacts can be utilized for developing efficient spintronic devices.

\begin{acknowledgments}
This work was supported by the Natural Science Foundation of Fujian Province of China (Grant No.2016J05163) and the Fundamental Research Funds for the Central Universities (Grant No. 20720160019). 
Zhenhua Wu was supported by the MOST of China (Grant No.2016YFA0202300).
Yang Yang and Cheng Li was supported by the National Basic Research Program of China (Grant No. 2013CB632103).
\end{acknowledgments}


\begin{thebibliography}{25}
\bibitem{note1} All the results of AlN can be refered to those of amorphous MgO very closely, since they have close parameters. But for crystalline MgO on Fe, these results may not be justified because the coherent tunneling through symmetry-matched bands is not includied in the calculations.
\bibitem{note2} The actual barrier height of TiO$_2$ could vary from 0.6 to -0.2 eV, depending on the growth methods or conditions, which means tuning $r_{b}^{*}$ by the barrier thickness is experimentally achievable.
\end{thebibliography}
\end{document}